\documentstyle[aps,prl,psfig,multicol]{revtex}

\begin{document}

\draft

\title{Anisotropic Spin Exchange in Pulsed Quantum Gates}

\author{N.E. Bonesteel and D. Stepanenko} 
\address
{Department of
Physics and National High Magnetic Field Laboratory, Florida State
University, Tallahassee, FL 32310} \author{D.P. DiVincenzo}
\address
{IBM Research Division, T.J. Watson Research Center, Yorktown
Heights, NY 10598}

\maketitle

\begin{abstract}
We show how to eliminate the first-order effects of the spin-orbit
interaction in the performance of a two-qubit quantum gate.  Our
procedure involves tailoring the time dependence of the coupling
between neighboring spins.  We derive an effective Hamiltonian which
permits a systematic analysis of this tailoring.  Time-symmetric
pulsing of the coupling automatically eliminates several undesirable
terms in this Hamiltonian.  Well chosen pulse shapes can produce an
effectively isotropic exchange gate, which can be used in universal
quantum computation with appropriate coding.
\end{abstract}



\vskip -.4in

\begin{multicols}{2}

The exchange interaction between spins is a promising physical
resource for constructing two-qubit quantum gates in quantum computers
\cite{loss,kane,burkard,hu,vrijen}. In the idealized case of vanishing
spin-orbit coupling, this interaction is isotropic, and any
Hamiltonian describing time-dependent exchange between two spin-1/2
qubits, $H_0(t) = J(t) {\bf S}_1 \cdot {\bf S}_2$, commutes with
itself at different times.  Thus, the resulting quantum gate depends
on $J(t)$ only through its time integral --- a convenient
simplification, particularly because, when carrying out quantum gates,
the exchange interaction should be pulsed adiabatically on time scales
longer than $\hbar/\Delta E$, where $\Delta E$ is a typical level
spacing associated with the internal degrees of freedom of the qubits
\cite{burkard}.  In addition, isotropic exchange alone has been shown
to be sufficient for universal quantum computation, provided the
logical qubits of the computer are properly encoded
\cite{bacon,divincenzo}.

Given the potential advantages of isotropic exchange for quantum
gates, it is important to understand the effect of the inevitable
anisotropic corrections due to spin-orbit coupling.  When these
corrections are included, the Hamiltonian describing time-dependent
exchange is
\begin{eqnarray}
H(t) = J(t) \left({\bf S}_1 \cdot {\bf S}_2 + {\cal A}(t)\right),
\label{pulse}
\end{eqnarray}
where
\begin{eqnarray}
{\cal A}(t) = {\mbox{\boldmath$\beta$}}(t) \cdot ({\bf S}_1 \times
{\bf S}_2) + {\bf S}_1\cdot {\rm \bf
I}\!{\mbox{\boldmath$\Gamma$}}(t)\cdot {\bf S}_2.
\end{eqnarray}
Here $\mbox{\boldmath$\beta$}(t)$ is the Dzyaloshinski-Moriya vector,
which is first order in spin-orbit coupling, and ${\rm \bf
I}\!{\mbox{\boldmath$\Gamma$}}(t)$ is a symmetric tensor which is
second order in spin-orbit coupling \cite{dizzymoriya}. Although these
corrections may be small, they will, in general, not be zero unless
forbidden by symmetry.  For example, Kavokin has recently estimated
that $\beta(t)$ can be as large as 0.01 for coupled quantum dots in
GaAs \cite{kavokin}.

In this Letter we construct the quantum gates produced by pulsing
$H(t)$.  This is nontrivial because $H(t)$ typically does not commute
with itself at different times.  We represent the resulting gates
using an effective Hamiltonian ${\overline H}(t)$, which we derive
perturbatively in powers of the spin-orbit coupling.  ${\overline
H}(t)$ is simple to work with because it {\it does} commute with
itself at different times.  As an application of this effective
Hamiltonian, we use it to tailor pulse forms that effectively
eliminate any first-order anisotropic corrections.

The quantum gate obtained by pulsing a particular $H(t)$ is found by
solving the time-dependent Schr\"odinger equation
$i\frac{d}{dt}|\Psi(t)\rangle = H(t) |\Psi(t)\rangle$ where
$|\Psi(t)\rangle$ is the state vector describing the two spin-1/2
qubits (here, and in what follows, $\hbar = 1$).  In general this
problem cannot be solved analytically.  However, since we expect
spin-orbit coupling to be small, it is natural to attempt a
perturbative solution in powers of $\mbox{\boldmath$\beta$}(t)$ and
${\rm \bf I}\!{\mbox{\boldmath$\Gamma$}}(t)$ .  To do this it is first
necessary to solve the unperturbed (${\cal A}(t) = 0$) problem
exactly.  This corresponds to pulsing the isotropic exchange
interaction, for which the unitary time evolution operator at time $t$
is
\begin{eqnarray}
U_0(t) &=& T \exp\left(-i\int_{-\infty}^t J(t^\prime) {\bf S}_1 \cdot
{\bf S}_2\; dt^\prime \right)
\nonumber\\ 
\nonumber\\ &=& \exp\left(-ix(t) {\bf
S}_1 \cdot {\bf S}_2 \right),
\label{u0}
\end{eqnarray}
where 
\begin{eqnarray}
x(t) = \int_{-\infty}^t J(t^\prime) dt^\prime.
\end{eqnarray} 
Here $T$ is the usual Dyson time ordering, and the second equality in
(\ref{u0}) follows from the fact that isotropic exchange commutes with
itself at different times.  The unperturbed quantum gate produced by a
full pulse is then $U_0(t\rightarrow\infty) = \exp(-i\lambda {\bf S}_1
\cdot {\bf S}_2)$ where $\lambda = \int_{-\infty}^{\infty} J(t) dt$ is
the pulse strength.  This is a well studied class of quantum gates
\cite{loss}.  For $\lambda = \pi$ the result is a simple swap, and for
$\lambda = \pi/2$ it is a `square root of swap' which, in conjunction
with single-qubit rotations, can be used to construct a controlled-not
gate \cite{loss}.

We now consider the effect of the anisotropic corrections ${\cal
A}(t)$. Given the evolution operator for the unperturbed system, it is
possible to recast the problem in the interaction picture by
introducing the state vector $|\Psi_I(t)\rangle \equiv U^\dagger_0(t)
|\Psi(t)\rangle$ which satisfies the Schr\"odinger equation
$i\frac{d}{dt} |\Psi_I(t) \rangle = J(t) {\cal A}_I(t)
|\Psi_I(t)\rangle$ where ${\cal A}_I(t) = U_0^\dagger(t) {\cal A}(t)
U_0(t)$. A formal expression for the unitary operator describing a
full pulse in this picture is then
\begin{eqnarray}
U_I = T \exp\left(-i\int_{-\infty}^\infty J(t) {\cal A}_I(t) dt\right).
\label{u}
\end{eqnarray}
Expanding the exponential in (\ref{u}) generates the standard
time-dependent perturbation theory expansion for $U_I$ in powers of
${\cal A}_I(t)$.  Returning to the Schr\"odinger picture, the unitary
operator describing the full quantum gate is $U = \exp(-i\lambda {\bf
S}_1\cdot{\bf S}_2) U_I$.

Rather than simply carrying out the perturbation expansion for $U_I$,
it is useful to parametrize the resulting quantum gate in terms of an
effective Hamiltonian of the form
\begin{eqnarray}
\overline H(t) = J(t)({\bf S}_1\cdot {\bf S}_2 + \overline
{\cal A}),
\end{eqnarray}
where the time dependence of $J(t)$ is the same as in $H(t)$, and
$\overline{\cal A}$ is independent of time.  Unlike $H(t)$, the
effective Hamiltonian $\overline H(t)$ commutes with itself at
different times. Thus, after a full pulse, ${\overline H}(t)$ yields
the quantum gate ${\overline U} = \exp\left(-i \lambda ({\bf S}_1
\cdot {\bf S}_2 + {\overline {\cal A}}) \right)$.  Our goal is then to
find the operator ${\overline {\cal A}}$ for which $\overline U$ is
equal to the quantum gate produced by a full pulse of $H(t)$.

Because $H(t)$ is traceless at all times $t$, the corresponding
unitary time evolution operator has determinant 1, i.e., $U \in$
SU[4]. Requiring that our effective Hamiltonian produce the same
quantum gate then implies that ${\overline {\cal A}}$ must also be a
traceless Hermitian operator.  The most general such operator acting
on the Hilbert space of two qubits can be written
\begin{eqnarray}
\overline{{\cal A}} = {\overline{\mbox{\boldmath$\beta$}}} \cdot ({\bf
S}_1 \times {\bf S}_2) &+& {\bf S}_1 \cdot
{\overline{{\rm \bf I}\!{\mbox{\boldmath$\Gamma$}}}}\cdot {\bf S}_2 \nonumber
\\ &+& \frac{\overline{{\mbox{\boldmath$\alpha$}}}}{2} \cdot ({\bf
S}_1 - {\bf S}_2) + \frac{\overline{{\mbox{\boldmath$\mu$}}}}{2} \cdot
({\bf S}_1 + {\bf S}_2),
\end{eqnarray}
where ${\overline{{\rm \bf I}\!{\mbox{\boldmath$\Gamma$}}}}$ is a
symmetric tensor.  This can be seen by noting that $\overline{\cal A}$
is indeed traceless and Hermitian, and has 15 independent real valued
parameters, the number of degrees of freedom for a 4$\times$4
traceless Hermitian matrix.

Before proceeding it is instructive to classify the terms in
${\overline {\cal A}}$ according to their symmetry properties under
inversion (${\bf S}_1 \leftrightarrow {\bf S}_2$) and time reversal
(${\bf S}_i \rightarrow -{\bf S}_i$).  Under inversion
$\overline{\mbox{\boldmath$\beta$}}$ and
$\overline{\mbox{\boldmath$\alpha$}}$ change sign, while
$\overline{{\rm \bf I}\!{\mbox{\boldmath$\Gamma$}}}$ and
$\overline{{\mbox{\boldmath$\mu$}}}$ do not.  Since
$\mbox{\boldmath$\beta$}(t)$ also changes sign under inversion this
implies that $\overline{\mbox{\boldmath$\beta$}}$ and
$\overline{\mbox{\boldmath$\alpha$}}$ are first order in spin-orbit
coupling, while $\overline{{\rm \bf I}\!{\mbox{\boldmath$\Gamma$}}}$
and $\overline{{\mbox{\boldmath$\mu$}}}$ are second order.  Under time
reversal $\overline{\mbox{\boldmath$\alpha$}}$ and
$\overline{{\mbox{\boldmath$\mu$}}}$ change sign, while
$\overline{\mbox{\boldmath$\beta$}}$ and $\overline{{\rm \bf
I}\!{\mbox{\boldmath$\Gamma$}}}$ are unaffected.  We therefore expect
that for time-reversal symmetric pulses, i.e., pulses for which $H(t_0
- t) = H(t)$ (where $t_0$ is the center of the pulse),
$\overline{{\mbox{\boldmath$\alpha$}}}$ and
$\overline{{\mbox{\boldmath$\mu$}}}$ will vanish.

To determine $\overline{\cal A}$ for a given pulse we note that the
requirement that $U = \overline U$ implies
\begin{eqnarray}
&&T\exp\left(-i\int_{-\infty}^\infty J(t) {\cal A}_I(t) dt\right)\nonumber\\
&&~~~~~~~~~~~~~~~~~~~~~~~~~~ = T
\exp\left(-i\int_{-\infty}^\infty J(t) {\overline{\cal A}}_I(t)
dt\right),
\label{pert}
\end{eqnarray}
where $\overline{\cal A}_I(t) = U^\dagger_0(t){\overline {\cal
A}}U_0(t)$.  Expanding both sides of (\ref{pert}) to a given order in
spin-orbit coupling and equating matrix elements yields a set of 15
independent equations.  These equations can then be solved for the
parameters in $\overline {\cal A}$ in terms of $J(t)$,
$\mbox{\boldmath$\beta$}(t)$ and ${\rm \bf
I}\!{\mbox{\boldmath$\Gamma$}}(t)$.

We have carried out this calculation to obtain the following
expressions valid to second order in spin-orbit coupling (i.e., second
order in $\mbox{\boldmath$\beta$}(t)$ and first order in ${\rm \bf
I}\!{\mbox{\boldmath$\Gamma$}}(t)$),
\begin{eqnarray}
{\overline { {\mbox{\boldmath$\alpha$}}}} = \frac{1}{2
\sin(\lambda/2)} \int_{-\infty}^{\infty} {\mbox{\boldmath$\beta$}}(t)
\sin \left( x(t) - \frac{\lambda}{2}\right) J(t) dt,
\label{alpha}
\end{eqnarray}
\begin{eqnarray}
{\overline {\mbox{\boldmath$\beta$}}} = \frac{1}{2 \sin(\lambda/2)}
\int_{-\infty}^{\infty} {\mbox{\boldmath$\beta$}}(t) \cos \left( x(t) -
\frac{\lambda}{2}\right) J(t) dt,
\label{beta}
\end{eqnarray}

\end{multicols}

\hrulefill

\begin{eqnarray}
{\overline { {\mbox{\boldmath$\mu$}}}} = \frac{1}{4\lambda}
\int_{-\infty}^{\infty} J(t_1)dt_1 \int_{-\infty}^{t_1} J(t_2) dt_2
\left( ({\mbox{\boldmath$\beta$}}(t_1) \times
{\mbox{\boldmath$\beta$}}(t_2)) \cos (x(t_1) - x(t_2)) +2({\overline {
{\mbox{\boldmath$\alpha$}}}} \times {\overline
{\mbox{\boldmath$\beta$}}}) \sin (x(t_1)-x(t_2))\right),
\label{mu}
\end{eqnarray}
and 
\begin{eqnarray}
\overline{{\rm I}\!{\Gamma}}_{ab} &=& \frac{1}{\lambda}
\int_{-\infty}^{\infty}  {\rm I}\!\Gamma_{ab}(t) J(t) dt
+\frac{1}{4\lambda}\int_{-\infty}^{\infty} J(t_1)dt_1
\int_{-\infty}^{t_1} J(t_2) dt_2 I_{ab}(t_1,t_2) \sin \left(x(t_1) -
x(t_2)\right),
\label{gamma}
\end{eqnarray}
where
\begin{eqnarray}
I_{ab}(t_1,t_2) =
2\left(\mbox{\boldmath$\beta$}(t_1)\cdot\mbox{\boldmath$\beta$}(t_2)
-{\overline {\beta}}^2
-{\overline {\alpha}}^2
\right) \delta_{ab} -\left(
\beta_a(t_1)\beta_b(t_2) +\beta_a(t_2)\beta_b(t_1) - 2 {\overline
\beta}_a {\overline \beta}_b - 2 {\overline \alpha}_a {\overline
\alpha}_b\right).
\end{eqnarray}

\begin{multicols}{2}

The criterion for the validity of these expressions is that $|\lambda
\overline{\mbox{\boldmath$\beta$}}|,
|\lambda\overline{\mbox{\boldmath$\alpha$}}| \ll 1$, where the factor
of $\lambda$ is included because it is the product
$\lambda\overline{\cal A}$ that enters the unitary operator $U$.  It
is then apparent that, for any finite $\mbox{\boldmath$\beta$}(t)$ and
${\rm \bf I}\!{\mbox{\boldmath$\Gamma$}}(t)$, our expansion breaks
down when $\lambda \rightarrow 2\pi n$ for $n = \pm 1, \pm 2,\cdots$,
because $\sin(\lambda/2) \rightarrow 0$ at these points.  However, for
$\lambda \rightarrow 0$, while $\overline{\mbox{\boldmath$\alpha$}}$
and $\overline{\mbox{\boldmath$\beta$}}$ may diverge, $\lambda
\overline{\mbox{\boldmath$\alpha$}}$ and
$\lambda\overline{\mbox{\boldmath$\beta$}}$ will always remain finite,
and so, provided $\mbox{\boldmath$\beta$}(t)$ and ${\rm \bf
I}\!{\mbox{\boldmath$\Gamma$}}(t)$ are small, our expansion remains
valid in this limit \cite{whyx(t)}.

As expected from symmetry considerations, we find that
$\overline{{\mbox{\boldmath$\beta$}}}$ and
$\overline{{\mbox{\boldmath$\alpha$}}}$ are first order in spin-orbit
coupling, while $\overline{{\rm \bf I}\!{\mbox{\boldmath$\Gamma$}}}$
and $\overline{{\mbox{\boldmath$\mu$}}}$ are second order.  It is also
readily verified that for a time-reversal symmetric pulse the
integrals (\ref{alpha}) and (\ref{mu}) for $\overline
{{\mbox{\boldmath$\alpha$}}}$ and $\overline{{\mbox{\boldmath$\mu$}}}$
vanish.  Thus these non time-reversal symmetric terms are only
generated by pulses that are themselves not time-reversal symmetric.

Given the possibility of using the exchange interaction alone to
perform universal quantum computation \cite{bacon,divincenzo}, which
depends crucially on the interaction being as close to isotropic as
possible, a natural questions arises: Is it possible to ameliorate the
effect of spin-orbit induced anisotropy on exchange-based quantum
gates?  We show below that the answer is yes --- by carefully shaping
pulses, it is possible to effectively eliminate the first-order
anisotropy terms leaving only a residual second-order anisotropy.

There are two first-order terms in $\overline H(t)$,
$\overline{{\mbox{\boldmath$\alpha$}}}$ and $\overline
{\mbox{\boldmath$\beta$}}$.  We have already seen how to eliminate
$\overline{{\mbox{\boldmath$\alpha$}}}$.  By choosing a time-reversal
symmetric pulse both $\overline{{\mbox{\boldmath$\alpha$}}}$ and
$\overline{{\mbox{\boldmath$\mu$}}}$ will vanish from $\overline
H(t)$.  Although $\overline{\mbox{\boldmath$\beta$}}$ cannot similarly
be eliminated, for appropriate pulse forms it can be effectively
eliminated by performing a local rotation in spin space.

Let ${\bf S}_2^\prime = {\rm \bf I}\!{\bf R} \cdot {\bf S}_2$ where
${\rm \bf I}\!{\bf R}$ is a rotation matrix constructed to eliminate
$\overline{\mbox{\boldmath$\beta$}}$ from $\overline H(t)$ so that
\begin{eqnarray} 
\overline H(t) = J(t)\left({\bf S}_1 \cdot {\bf S}^\prime_2 + {\bf
S}_1 \cdot {\overline{{\rm \bf I}\!{\mbox{\boldmath$\Gamma$}}}}^\prime
\cdot {\bf S}_2^\prime\right),
\label{rotated}
\end{eqnarray}
where ${\overline{{\rm \bf I}\!{\mbox{\boldmath$\Gamma$}}}}^\prime$ is
a symmetric tensor.  The precise form of this rotation depends on both
$\overline{\mbox{\boldmath$\beta$}}$ and $\overline{{\rm \bf
I}\!{\mbox{\boldmath$\Gamma$}}}$ and cannot be expressed simply.
However, up to second order in
$\overline{\mbox{\boldmath$\beta$}}$, it is given by
\begin{eqnarray}
{\rm I\!R}_{ab} = \delta_{ab} +
\sum_c\epsilon_{abc}{\overline\beta}_c - ({\overline\beta}^2
\delta_{ab} -{\overline\beta}_a{\overline\beta}_b)/2 +
O({\overline\beta}^3),
\label{rmatrix}
\end{eqnarray}
and this is sufficient for our purpose of eliminating first-order
anisotropy. Using (\ref{rmatrix}) one finds the residual anisotropy in
(\ref{rotated}) is, up to second order in
$\overline{\mbox{\boldmath$\beta$}}$,
\begin{eqnarray}
\overline{{\rm I}\!\Gamma}_{ab}^\prime = \overline{ {\rm
I}\!\Gamma}_{ab} + ({\overline\beta}^2 \delta_{ab} -
{\overline\beta}_a {\overline\beta}_b)/2 + O({\overline\beta}^4).
\label{gammaprime}
\end{eqnarray}
Thus, in this rotated coordinate system the first-order anisotropy
vanishes and all corrections to the isotropic exchange interaction are
second order in spin-orbit coupling.

The ability to eliminate $\overline{\mbox{\boldmath$\beta$}}$ from
$\overline H(t)$ by simply rotating one qubit with respect to the
other indicates a procedure for eliminating the first-order effects of
spin-orbit coupling in any quantum computer that uses tunable exchange
for quantum gates.  Suppose that symmetric pulses are used, so that
$\overline{\mbox{\boldmath$\alpha$}} = 0$, and pulse forms are chosen
so that $\overline{\mbox{\boldmath$\beta$}}$ is the same for all pulse
strengths $\lambda$.  Then, if the qubits in the computer form a linear
array, or any arrangement for which there are no closed loops of
qubits connected by two-qubit gates, it will be possible to define a
{\it local spin-space coordinate system} in which the effective
interaction between any two neighboring qubits has the form
(\ref{rotated}).  While this procedure does not completely eliminate
the anisotropy, it does reduce it from an effect that is first order
in spin-orbit coupling to one that is second order.

To demonstrate how (\ref{beta}) can be used to tailor pulse shapes
that lead to the same $\overline{\mbox{\boldmath$\beta$}}$ for all
pulse strengths $\lambda$, consider the family of pulses
\begin{eqnarray}
J(t;\lambda) = J_0(\lambda)~{\rm sech}^2(2t/\tau(\lambda)),
\label{sech}
\end{eqnarray}
where $J_0(\lambda)$ and $\tau(\lambda)$ are, respectively, the pulse
height and width, and the pulse strength is $\lambda =
\int_{-\infty}^\infty J(t;\lambda) = J_0(\lambda) \tau(\lambda)$.  To
evaluate (\ref{beta}) it is also necessary to know the time dependence
of $\mbox{\boldmath$\beta$}(t)$.  Determining the precise form of this
dependence will require a detailed microscopic study of the specific
realization of the exchange interaction being considered. Here we
take, as the simplest possible illustrative model, a linear dependence
on $J(t;\lambda)$,
\begin{eqnarray}
{\mbox{\boldmath$\beta$}}(t) = {\mbox{\boldmath$\beta$}}_1
J(t;\lambda),
\label{linear}
\end{eqnarray}
for which the integral (\ref{beta}) can be performed analytically,
with the result
\begin{eqnarray}
{\overline {\mbox{\boldmath$\beta$}}} 
&=& {\mbox{\boldmath$\beta$}}_1 \frac{4 J_0(\lambda)}{\lambda^2}
\left(2 - \lambda \cot(\lambda/2)\right).
\label{betalambda}
\end{eqnarray}
Also, because these pulses are time-reversal symmetric, (\ref{alpha})
gives $\overline{\mbox{\boldmath$\alpha$}} = 0$.

Equation (\ref{betalambda}) can be used to exploit the freedom to
choose $J_0(\lambda)$ and $\tau(\lambda)$, while keeping
$J_0(\lambda)\tau(\lambda) = \lambda$, to shape pulses that keep
$\overline {\mbox{\boldmath$\beta$}}$ fixed for different pulse
strengths.  For example, if the pulse parameters for $\lambda = \pi$
(swap) are fixed to be $J_0(\pi)$ and $\tau(\pi)$, then, for general
$\lambda$, one should take
\begin{eqnarray}
J_0(\lambda) &=& J_0(\pi) \frac{2\lambda^2}{\pi^2} \frac{1}{2 -
\lambda \cot(\lambda/2)},
\label{j0}
\end{eqnarray}
and
\begin{eqnarray}
\tau(\lambda) = \tau(\pi) \frac{\pi}{2\lambda} (2 -\lambda
\cot(\lambda/2)).
\label{tau}
\end{eqnarray}

\begin{figure}[t]
\centerline{\psfig{figure=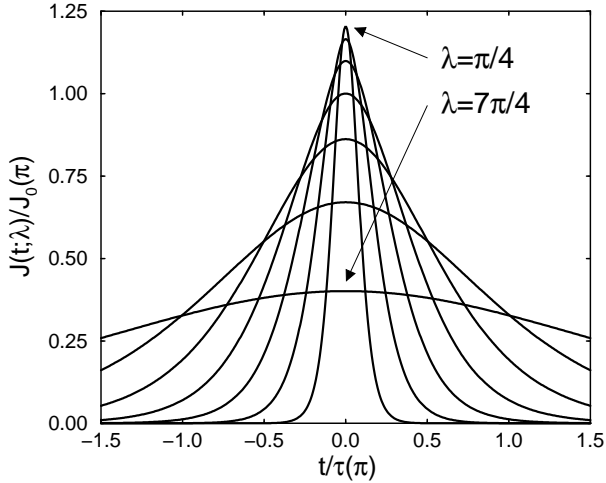,height=2.9in,angle=270}}
\vskip .15in
\caption{Pulse forms tailored to produce the same
{${\overline{\beta}}$} for different pulse strengths $\lambda$ for the
example described in the text.  Of the pulses shown, the narrowest
with the highest peak is for $\lambda = \pi/4$. $\lambda$ then
increases in increments of $\pi/4$ as the peak height decreases until,
for the widest pulse with the lowest peak, $\lambda = 7 \pi/4$.  As
$\lambda \rightarrow 2\pi$ the pulse height goes to zero.}
\label{pulsefig}
\end{figure}

These pulse forms are shown in Fig.~\ref{pulsefig} for various values
of pulse strength $\lambda$.  Note that as $\lambda$ increases, the
pulse height {\it decreases}.  This is because
$\overline{\mbox{\boldmath$\beta$}}$ becomes increasingly sensitive to
$\mbox{\boldmath$\beta$}(t)$ with increasing $\lambda$ until, in the
limit $\lambda \rightarrow 2\pi$, the pulse height must go to zero if
$\overline{\mbox{\boldmath$\beta$}}$ is to be kept constant.  Although
our perturbation expansion for $\overline{\cal A}$ breaks down as
$\lambda \rightarrow 2\pi$, for this example the pulse heights are
chosen so that the parameters in $\overline{\cal A}$ remain small, and
we are always within the perturbative regime. The pulse forms defined
by (\ref{j0}) and (\ref{tau}) are therefore valid, even in this
singular limit.  Of course, in practice, pulses near $\lambda = 2\pi$
will be problematic because of the diverging pulse length.

\vskip -.2in

Once the first-order corrections to $\overline H(t)$ are eliminated,
the residual second-order anisotropy can be found by first evaluating
(\ref{gamma}) and then performing the local rotation to eliminate
$\overline{\mbox{\boldmath$\beta$}}$.  As a specific example, consider
the special case for which the form of the pulsed Hamiltonian is
\begin{eqnarray}
H(t) = J(t)~{\bf S}_1 \cdot {\rm \bf I}\!{\bf R}(t) \cdot {\bf S}_2,
\label{rotatedexchange}
\end{eqnarray}
where ${\rm \bf I}\!{\bf R}(t)$ is a time-dependent rotation matrix.
Such rotated exchange is, in fact, precisely the form of anisotropy
found microscopically when spin-orbit corrections are included in the
usual Hubbard model treatment of superexchange
\cite{shekhtman,bonesteel}.  It has also been suggested that this form
is appropriate for localized electrons in semiconductors
\cite{kavokin}. In the present context (\ref{rotatedexchange}) is of
interest because if the rotation matrix ${\rm \bf I}\!{\bf R}(t)$ were
independent of time, our local rotation scheme would eliminate
anisotropy to {\it all} orders, rather than just to first order in
spin-orbit coupling.  It is therefore natural to ask to what degree
the fact that ${\rm \bf I}\!{\bf R}(t)$ depends on time spoils this
hidden symmetry.

For the particular form of anisotropic exchange in
(\ref{rotatedexchange}), the symmetric anisotropy term is, to second
order in $\mbox{\boldmath$\beta$}(t)$,
\begin{eqnarray}
{\rm I}\!\Gamma_{ab}(t) = -\left(\beta(t)^2 \delta_{ab}
-\beta_a(t)\beta_b(t)\right)/2 + O(\beta(t)^4).
\end{eqnarray}
For this ${\rm \bf I}\!{\mbox{\boldmath$\Gamma$}}(t)$, if we continue
to take the pulse form (\ref{sech}) and $\mbox{\boldmath$\beta$}(t)$
from (\ref{linear}) then the expression (\ref{gamma}) can be evaluated
analytically.  After performing the local rotation to eliminate
$\overline{\mbox{\boldmath$\beta$}}$ we find, using
(\ref{gammaprime}), that the residual anisotropy in ${\overline H}(t)$
is
\begin{eqnarray}
\overline{{\rm I}\!\Gamma}^\prime_{ab} &=& \frac{8
J_0(\lambda)^2}{3\lambda^4} \left(\lambda^2 +
6\lambda\cot(\lambda/2)-12 \right)\nonumber\\
&&~~~~~~~~~~~~~~~~~~~~\times (\beta_1^2 \delta_{ab} - {\beta_1}_a
{\beta_1}_b) + O(\beta_1^4).
\end{eqnarray}
Thus even for the rotated exchange (\ref{rotatedexchange}), if the
rotation depends on time we are still left with residual second-order
anisotropy after a pulse.

To summarize, we have studied the effects of anisotropic corrections
due to spin-orbit coupling on quantum gates produced by pulsing the
exchange interaction between two spin-1/2 qubits.  These quantum gates
are parametrized by an effective Hamiltonian that commutes with itself
at different times and produces the same quantum gate as a given
pulse. Expressions for the various parameters in this effective
Hamiltonian are obtained perturbatively in powers of spin-orbit
coupling and used to shape pulses that effectively eliminate
first-order spin-orbit corrections to quantum gates.  The ability to
reduce spin-orbit effects from first order to second order should be
useful for any quantum computing scheme which relies on isotropic
exchange.

N.E.B.\ and D.S.\ acknowledge support from U.S. DOE Grant No.\
DE-FG02-97ER45639 and the NSF.  D.P.D.\ is grateful for support from
the National Security Agency and the Advanced Research and Development
Activity through Army Research Office contract DAAG55-98-C-0041.

\end{multicols}


\begin{references}
\bibitem{whyx(t)} $\overline\alpha$ and $\overline\beta$ will only
diverge as $\lambda \rightarrow 0$ if $J(t)$ changes sign.  If $J(t)$
is always positive (or always negative) our expressions for the
parameters in $\overline{\cal A}$ can be simplified by changing
integration variables from $t$ to $x = x(t)$ and using the fact that
$dx = J(t) dt$.  It is then easily seen that $\overline{\alpha}$ and
$\overline\beta$ remain finite as $\lambda \rightarrow 0$ for these
pulses.
\end{references}
\end{document}